\documentclass[preprintnumbers, floatfix, showkeys, preprintnumbers, letterpaper, twocolumn, superscriptaddress,nofootinbib]{revtex4-2}
\pdfoutput=1 
\usepackage{graphicx}
\usepackage{amsmath}
\usepackage{amssymb}
\usepackage{subfigure}
\usepackage{url}
\usepackage{hyperref}

\usepackage{orcidlink}

\usepackage{xcolor}
\usepackage{color}
\usepackage{mathrsfs}
\usepackage{calrsfs}
\usepackage{amsfonts}
\usepackage{tabularx}
\usepackage{latexsym}
\usepackage{ragged2e}
\usepackage{epsfig}
\usepackage{textcomp}
\usepackage{float}

\usepackage{caption}
\DeclareCaptionJustification{justified}{\leftskip=0pt \rightskip=0pt \parfillskip=0pt plus 1fil}
\definecolor{vividviolet}{rgb}{0.62, 0.0, 1.0}
\definecolor{amaranth}{rgb}{0.9, 0.17, 0.31}
\definecolor{palatinateblue}{rgb}{0.15, 0.23, 0.89}
\definecolor{brightpink}{rgb}{1.0, 0.0, 0.5}
\definecolor{cornflowerblue}{rgb}{0.39, 0.58, 0.93}
\definecolor{deepcarminepink}{rgb}{0.94, 0.19, 0.22}
\definecolor{radicalred}{rgb}{1.0, 0.21, 0.37}

\hypersetup{ linktoc=all,
	colorlinks, linkcolor={palatinateblue},
	citecolor={brightpink}, urlcolor={amaranth}
}
\newcommand{\changeurlcolor}[1]{\hypersetup{urlcolor=#1}}
\graphicspath{{Images/}}

\def\sideremark#1{\ifvmode\leavevmode\fi\vadjust{\vbox to0pt{\vss
			\hbox to 0pt{\hskip\hsize\hskip1em
				\vbox{\hsize1.3cm\tiny\raggedright\pretolerance10000
					\noindent #1\hfill}\hss}\vbox to8pt{\vfil}\vss}}}%
\def\beq{\begin{equation}}
\def\eeq{\end{equation}}

\begin{document}
\title{Hawking Temperature and the Inverse-Radius Scale of the Horizon}

\author{Michael R.R. Good\orcidlink{0000-0002-0460-1941}}
\email{michael.good@nu.edu.kz}
	\affiliation{Department of Physics \& Energetic Cosmos Laboratory, Nazarbayev University, Astana 010000, Qazaqstan}
	\affiliation{Leung Center for Cosmology and Particle Astrophysics,
National Taiwan University, Taipei 10617, Taiwan}

\author{Yen Chin \surname{Ong}\orcidlink{0000-0002-3944-1693}}
\email{ycong@yzu.edu.cn}
\affiliation{Center for Gravitation and Cosmology, College of Physical Science and Technology,\\ Yangzhou University,
 Yangzhou, 225002, China}
\affiliation{Shanghai Frontier Science Center for Gravitational Wave Detection, School of Aeronautics and Astronautics,\\ Shanghai Jiao Tong University, Shanghai 200240, China}

\begin{abstract}
The Hawking temperature of a Schwarzschild black hole can be heuristically derived by identifying the temperature with the inverse radius of the horizon up to a multiplicative constant. This does not work for more general black holes such as the Kerr and Reissner-Nordstr\"om solutions. Expounding on the details of how it fails to work nevertheless uncovers some connections with the ``spring constant'' of black holes and with black hole thermodynamics. 
\end{abstract} 
\maketitle
\section{Introduction: The Inverse Radius Scale of a Black Hole Horizon}
In the Generalized Uncertainty Principle (GUP) literature, one can often find heuristic derivations of the Hawking temperature. The canonical example is for an asymptotically flat Schwarzschild black hole (before any GUP correction), whose Hawking temperature is known to be
\begin{equation}
T=\frac{\hbar c^3}{8\pi G k_B M}.
\end{equation}
The derivation is based on the Heisenberg uncertainty principle \cite{0106080}: $\Delta x\Delta p \geqslant \hbar/2$. One identifies the characteristic temperature with $T \sim c\Delta p/k_B$, and the uncertainty in the position $\Delta x \sim r_+$ is seen as the horizon scale (see also \cite{1808.05121}). That is,
\begin{equation}
T \sim \frac{\hbar c}{2 k_B \Delta x} = \frac{\hbar c^3}{4 G k_B M}.
\end{equation}
A calibration factor of $1/2\pi$ is needed to obtain the correct temperature. In other words, the Hawking temperature can be obtained as
\begin{equation}\label{pre}
T = \frac{1}{4\pi}\left(\frac{1}{r_+}\right),
\end{equation}
so that it is, up to a constant prefactor $1/4\pi$, the horizon's inverse-radius scale (IRS). 
In Eq.~(\ref{pre}), we have switched to the commonly used units in which $\hbar=G=c=k_B=1$, which will be employed hereinafter.

Besides dimensional analysis, this heuristic derivation works because the identification $\Delta x \sim r_+$ corresponds to the fact that Hawking particles are being emitted at some distance away from the black hole, in the range of the so-called ``quantum atmosphere'' \cite{1511.08221,1607.02510,1701.06161,2003.10429} (though see \cite{2405.08167} for subtleties). However, one does not expect this to work when the black hole is sufficiently charged since $T \to 0$ in the extremal limit, but $1/r_+$ remains finite. 

One could conclude that the inverse radius scale is useless for more general black holes and that the heuristic derivation only works for Schwarzschild, perhaps due to a coincidence. Yet, as we shall demonstrate in this short article, although Eq.~(\ref{pre}) is not the Hawking temperature, it is nevertheless an interesting quantity. We will look at two black hole solutions in general relativity: the electrically charged Reissner-Nordstr\"om case and the rotating Kerr case. (The Kerr case turns out to be simpler.)

\section{The Reissner-Nordsrt\"om Case}
The Reissner-Nordstr\"om black hole has a temperature
\begin{equation}
T_\text{RN} = \frac{\sqrt{M^2-Q^2}}{2\pi r_+^2}=\frac{1}{8\pi M}-\frac{x^4}{128\pi M} + \cdots,
\end{equation} 
where $x:=Q/M$, and we assume $M\geqslant Q$, without loss of generality (WLOG). 
Here, we have expanded the temperature as a Taylor series of $x$ around $x=0$.
On the other hand, with the horizon radius $r_+=M+\sqrt{M^2-Q^2}$, the prescription Eq.~(\ref{pre}) would instead give
\begin{equation}\label{IRSRN}
T_\text{IRS} =\frac{1}{8\pi M}+\frac{x^2}{32\pi M} + \frac{x^4}{64\pi M}+ \cdots.
\end{equation}
We immediately note that $T_\text{IRS}$ contains an $x^2$ term, but $T_\text{RN}$ does not. In addition, the subleading terms (i.e., the deviation from Schwarzschild temperature) in $T_\text{IRS}$ are of the opposite sign compared to $T_\text{RN}$. This is expected: for the same mass, the horizon radius $r_+$ is smaller for the Reissner-Nordstr\"om black hole compared to the Schwarzschild counterpart, so the ``temperature'' associated with the IRS is higher; on the other hand, we know that adding electrical charges lowers the temperature of the black hole. It is, therefore, clear that the heuristic method cannot work. This does not mean that $T_\text{IRS}$ is useless. On the contrary, we claim it encodes certain information about black hole thermodynamics. 

Having brought out the issues with series expansion, we may now turn to the full analysis.
Instead of Eq.~(\ref{pre}), the Reissner-Nordstr\"om black hole temperature satisfies
\begin{equation}
T = \frac{1}{2\pi}\left(\frac{1}{2r_+}+\frac{C(x)}{r_+}\right),
\end{equation}
where the ``correction term'' is given by
\begin{equation}\label{C}
C(x):=\frac{\sqrt{1-x^2}}{1+\sqrt{1-x^2}}-\frac{1}{2}.
\end{equation}
Equivalently, in the form reminiscent of the uncertainty principle (in which $2\pi$ is the calibration factor),
\begin{equation}
2\pi r_+ T = \frac{1}{2} + C(x).
\end{equation}
The case $C(x=0)=0$ reduces to the Schwarzschild one. 

We note that
\begin{equation}
C(x)= -\frac{1}{4} \Longleftrightarrow x = \sqrt{\frac{8}{9}}\approx 0.9428.
\end{equation}
The magnitude $1/4$ may be important as it relates to the ``maximum force conjecture'' in general relativity\footnote{The concept of ``force'' and what kind of ``force'' is subject to the conjecture remains a debate, but this concept is not necessary for our work; our focus here is primarily on the thermodynamics of black holes. We only mention this for readers who may be interested in this issue.} \cite{0210109,1408.1820,0607090,724159}. Indeed, $x=\sqrt{8/9}$ was found to correspond to the value at which the ``thermodynamic force'' of a Reissner-Nordstr\"om (as well as Kerr) black hole attains the maximum force value \cite{2309.04110}. The thermodynamic force satisfies
\begin{equation}
F_\text{therm}(x)=\frac{\partial M}{\partial r_+}=\frac{\sqrt{1-x^2}}{1+\sqrt{1-x^2}},
\end{equation}
where $M$ is the mass function $M(Q,r_+)=(Q^2+r_+^2)/2r_+$.
Therefore, we see that Eq.~(\ref{C}) can be re-stated as
\begin{equation}
C(x)=F_\text{therm}(x)-\frac{1}{2}.
\end{equation}

Next, we consider the error ratio between the actual Hawking temperature $T$ and $T_\text{IRS}$, 
\begin{equation}
R(x):=\frac{T_\text{IRS}-T}{T}.
\end{equation}
For the Reissner-Nordstr\"om case, we have
\begin{equation}
R(x)=\frac{T_\text{IRS}-T_\text{RN}}{T_\text{RN}}=\frac{1}{2}\left(\frac{1-\sqrt{1-x^2}}{\sqrt{1-x^2}}\right).
\end{equation}
This is a monotonically increasing function in $x \in [0,1)$ that diverges in the extremal limit $x \to 1$. We can check that 
\begin{equation}
R(x)= 1 \Longleftrightarrow x = \sqrt{\frac{8}{9}}.
\end{equation}
That is, for $x > \sqrt{8/9}$, the ``error'' is greater than 100\%.
We leave the interpretation of this result to the Discussion section.

\section{The Kerr Case}

For the Kerr black hole, its Hawking temperature is 
\begin{flalign}
T_\text{Kerr}&=\frac{1}{4\pi M}\frac{\sqrt{M^2-a^2}}{M+\sqrt{M^2-a^2}} \\&= \frac{1}{8\pi M} - \frac{a^2}{32 \pi M^3} - \frac{a^4}{64\pi M^5} - \cdots.
\end{flalign}
The IRS gives 
\begin{equation}
T_\text{IRS}= \frac{1}{8\pi M} + \frac{a^2}{32 \pi M^3} + \frac{a^4}{64\pi M^5} + \cdots,
\end{equation}
the correction terms of which are \emph{exactly the same but opposite in sign} compared to $T_\text{Kerr}$. 
That is to say, 
\begin{equation}\label{sumofT}
T_\text{IRS} + T_\text{Kerr} = 2T_\text{Sch},
\end{equation}
where $T_\text{Sch}$ denotes the temperature of a Schwarzschild black hole of the same mass. Instead of comparing term by term the series expansion of $T_\text{IRS}$ and $T_\text{Kerr}$, we can instead prove Eq.~(\ref{sumofT}) directly. This is done by some elementary algebraic manipulation, after which Eq.~(\ref{sumofT}) can be re-written in a form already obtained in our previous work from 10 years ago that expresses the Hawking temperature of Kerr black hole as a kind of deviation from the Schwarzschild case \cite{1412.5432}:
\begin{equation}
T_\text{Kerr} = \frac{1}{2\pi} (\kappa_\text{Sch}-k),
\end{equation}
where $\kappa_\text{Sch}$ is the surface gravity of the Schwarzschild black hole of the same mass, and $k$ is the ``spring constant''\footnote{The name ``spring constant'' was coined because in (3+1)-dimensions, $k=M\Omega_+^2$, where $\Omega_+$ is the angular velocity of the horizon (therefore $k$ is only a true constant if $M$ and $a$ are fixed) \cite{1412.5432}. This reminds us of the spring constant $k=m\omega^2$ in elementary physics. In addition, the quantity $F:=k r_+$, naively defined analogously according to the Hookean law for a spring, is bounded above $F\leqslant 1/4$, with equality attained at extremality -- which again is consistent with the maximum force conjecture \cite{1412.5432} (the subtlety being $F$ is not a force -- see \cite{2108.13435} for discussions.). We can similarly obtain $k=1/2r_\text{Sch}-1/2r_+ - C/r_+$ for Reissner-Nordstr\"om black holes from the relation $T=(1/2\pi)(\kappa_{\text{Sch}}-k)$.} defined in \cite{1412.5432}. 

Denote $x:=a/M$. Again, we assume $a \leqslant M$, WLOG.
We can obtain with similar calculations the ``correction'' term, 
\begin{equation}
C(x)=\frac{\sqrt{1-x^2}}{2}-\frac{1}{2}.
\end{equation}
In this case, the first term does not correspond to $F_\text{therm}$. Nevertheless, the error ratio 
\begin{equation}
R(x)=\frac{T_\text{IRS}-T_\text{Kerr}}{T_\text{Kerr}}=\frac{1-\sqrt{1-x^2}}{\sqrt{1-x^2}}
\end{equation}
still reveals some interesting properties.
For example, we note that
\begin{equation}\label{2factor}
R^\text{Kerr}(x) =  2R^\text{RN}(x).
\end{equation}
This is unexpected given how complicated the Kerr geometry is compared to the relatively simpler Reissner-Nordstr\"om spacetime. Specifically, the Kerr temperature has $M(M+\sqrt{M^2-a^2})$ in the denominator, c.f. the more straightforward, more symmetric $(M+\sqrt{M^2-Q^2})^2$ form for the Reissner-Nordstr\"om case. Perhaps of relevance is the fact that the ratio of $T_\text{RN}(x)/T_\text{Kerr}(x)$ in the extremal limit is 2, although both temperatures are zero in that limit.

In addition, for the Kerr case, we have
\begin{equation}
R(x)= 1 \Longleftrightarrow x = \frac{\sqrt{3}}{2}\approx 0.866.
\end{equation}

\begin{figure}[h!] 
\centering
\includegraphics[width=.5\textwidth]{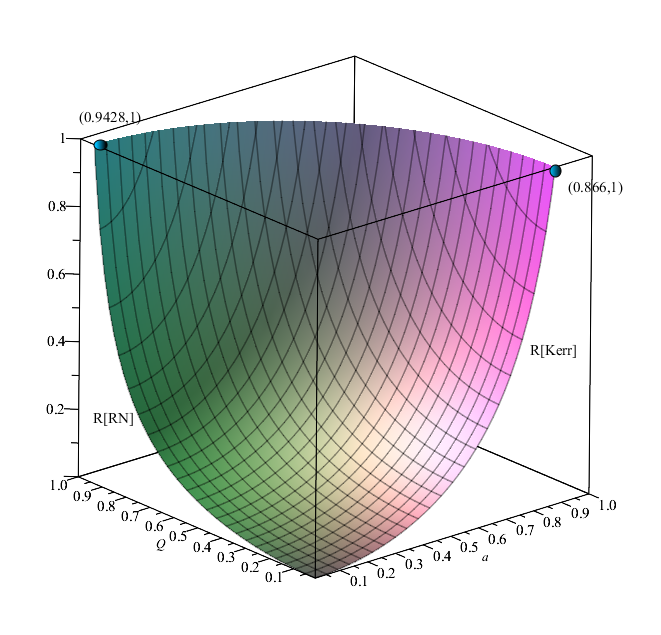}
\caption{The error ratio for the Kerr-Newman black hole temperature, where we have only shown the range between $0$ and $1$. The ``right-facing'' curve is $R^\text{Kerr}$, and the ``left-facing'' curve is $R^\text{RN}$. The points they attain unity is $(\sqrt{3}/2,1)$ and $(\sqrt{8/9},1)$, respectively. \label{R}}
\end{figure}

One might be worried about whether one could compare $R^\text{Kerr}$ and $R^\text{RN}$ in such a manner, since the Reissner-Nordstr\"om metric employs polar spherical coordinates, whereas the Kerr metric uses oblate spheroidal coordinates where $r$ is defined differently and given a different physical meaning. Still, $R$ is a quantity defined solely in terms of the Hawking temperatures and $T_\text{IRS}$, the latter can also be regarded as the temperature of a black hole of the same $r_h$, but one which is not rotating and has no charge. In this way, the quantities are physically defined to the same observer class (static observers at asymptotic infinity). One could also simply compute $R$ for Kerr-Newmann geometry and compute the $R$ that way, then take the limit $a \to 0$ and $Q \to 0$ separately, thus ensuring one works in the same coordinates. The error ratio for the Kerr-Newman black hole is shown in Fig.(\ref{R}). Note that of course the relation Eq.(\ref{2factor}) can be obtained directly from the algebraic expression, the mystery is not the factor of 2 \emph{per se}, but rather why $x=1$ corresponds to thermodynamically relevant points for both the Kerr and Reissner-Nordstr\"om case. Whether the more general case with both $Q\neq 0, a\neq 0$ but $R=1$ has any thermodynamic relevance would require additional investigations. Even if it does not, the fact that when only $Q$ or $a$ are present is thermodynamically relevant may have a deeper underlying physics.

\section{Discussion: The Physics of the Error Ratio}
Given our findings above, 
in terms of the spring constant $k$ and the quantity $F:=kr_+$, we can write
\begin{equation}\notag
2\pi r_+ T = 
\begin{cases}
\frac{1}{2} - 2F,  ~~~~~~~~~~~~~~~~~~ \text{for Kerr};\\
\frac{1}{2} - \left[F+\frac{1}{2} - \frac{r_+}{2r_\text{Sch}}\right], \text{for Reissner-Nordstr\"om},
\end{cases}
\end{equation}
where $r_\text{Sch}$ denotes the horizon radius of the Schwarzschild black hole of the same mass.
The cleaner and simpler expression for the Kerr case reflects that the subleading terms of its Hawking temperature have the same magnitude (but a different sign) as $T_\text{IRS}$. This is contrary to our expectation that, usually, in many circumstances, the Reissner-Nordstr\"om case gives rise to simpler expressions due to spherical symmetry. 

The error ratio $R(x)$ has a physical interpretation. To understand its significance, we first recall that black holes are quite peculiar compared to ``ordinary'' black bodies in the sense that the Hawking wavelength is larger than the horizon size. For a Schwarzschild black hole, $\lambda_T = 8\pi^2 r_+$  \cite{1511.08221}. Still, the coefficient is ``only'' $\mathrm{O}(10)$ or so of the Schwarzschild radius. In the charged or rotating case, as the black hole tends towards extremality, $\lambda_T$ becomes arbitrarily long compared to the horizon size. Naively, but essentially, one can treat $\lambda_T$ as the de Broglie's wavelength, and therefore, it gives the scale from which the Hawking particle can be emitted (i.e., the size of the quantum atmosphere). As the amount of charge, or angular momentum, increases, particle emissions can occur further from the horizon \cite{2003.10429}. The ratio $R(x)$ essentially measures the deviation between $1/\lambda_T$ and $1/r_+$, modulo some constant prefactor for each, or equivalently, the deviation between the quantum wavelength $\lambda_T$ and the classical scale $r_+$. 

It is, therefore, interesting that for Reissner-Nordstr\"om black holes $R(x)=1$ when $x=\sqrt{8/9}\approx 0.9428$, i.e., the charge-to-mass ratio at which it was argued that the black hole becomes ``quantum dominated'' \cite{2309.04110}, evidenced by various curious behaviors in the properties of its Hawking radiation \cite{2003.10429,2301.12319,0805.1876,9607048}. For the Kerr case, $x=\sqrt{8/9}$ does not seem to correspond to any peculiarity in the Hawking radiation, except perhaps that for a black hole that rotates faster than this (approximate) value of $x=a/M$ the superradiance effect dominates over the Hawking one \cite{2306.17423}. \emph{Classically}, however, the equatorial innermost stable circular orbit (ISCO) coincides with the ergosphere when $x=\sqrt{8/9}$. In terms of $R(x)$, for Kerr black holes, $x=\sqrt{8/9}$ corresponds to $R(x)=2$, instead of $R(x)=1$. The latter, however, as we have seen, equals $\sqrt{3}/2$. This value is also classically relevant for Kerr black holes: it is the largest value of $x$ that the corresponding black hole horizon geometry can be isometrically embedded in $\Bbb{R}^3$, since for $x>\sqrt{3}/2$ the Gaussian curvature close to the poles becomes negative \cite{0706.0622}. Due to the relation $R^\text{Kerr}(x) =  2R^\text{RN}(x)$, a Reissner-Nordstr\"om black hole with $x=\sqrt{3}/2$ corresponds to $R=1/2$. One notes that this value is also significant since it corresponds to the well-known Davies point, beyond which the specific heat $C := dM/dT$ for a Reissner-Nordström black hole with a fixed charge becomes positive \cite{davies}. On the other hand, smooth embedding for the $(t,r)$-part of Reissner-Nordstr\"om geometry into (2+1)-dimensional Minkowski spacetime requires $Q/M < \sqrt{8/9}$ \cite{0305102}.

Thus, we have shown that the quantity $T_\text{IRS}$, although \emph{not} equal to the Hawking temperature in the more general case\footnote{This is, by itself, an important point. The GUP literature is full of heuristic ``derivations'' of the Hawking temperature of various black holes. However, as we have seen, this may not give the correct result once the geometry is sufficiently complicated.}, nevertheless encodes some physics related to black hole geometry and thermodynamics. Specifically, we have seen that the error ratio $R(x)$ for Kerr black holes is {exactly twice} that of Reissner-Nordstr\"om black hole. This can be derived directly from the Kerr-Newman expression. Its consequence is that when the error ratios attain unity for both the $Q=0$ and $a=0$ cases, the black holes have special thermodynamic properties—understanding why may help us to understand black hole thermodynamics further.

It should be noted that the Hawking temperature, as well as $T_\text{IRS}$, has a one-to-one correspondence with the classical surface gravity $\kappa$ via the equation
$2\pi c k_BT={\hbar \kappa}$,
where we have explicitly restored the physical constants for clarity. Therefore, some of the peculiarities can be revealed by looking at the \emph{classical} quantity $\kappa$ for some classes of observers, as was done in \cite{2301.12319}. However, 
quantum spectrum analysis reveals interesting phenomena distinct from classical analysis alone (for example, the semi-classically divergent spectrum of Hawking radiation associated with the inner horizon is distinct from the classical mass inflation \cite{2301.12319}). Thus, whether the interesting values of $Q/M$ or $a/M$ truly correspond to quantum or classical behaviors would require more studies. 

Finally, for completeness, we cannot help but mention another curiosity about the Kerr black hole. Eq.~(\ref{sumofT}) is equivalent to saying that $T_\text{Sch}$ is the arithmetic mean of $T_\text{IRS}$ and $T_\text{Kerr}$. It is, therefore, intriguing to note that Kerr black holes also satisfy the following relation \cite{OK},
\begin{equation}
\frac{1}{T_\text{Kerr}} + \frac{1}{T_-} = \frac{1}{T_\text{Sch}},
\end{equation}
where $T_-$ is the inner horizon temperature. That is, $T_\text{Sch}$ is half the harmonic mean of $T_\text{Kerr}$ and $T_-$.

\begin{acknowledgments}
Funding comes partly from the FY2024-SGP-1-STMM Faculty Development Competitive Research Grant (FDCRGP) no.201223FD8824 and SSH20224004 at Nazarbayev University in Qazaqstan. 
\end{acknowledgments}


\end{document}